\documentstyle[aps,prl,amssymb,twocolumn,epsfig]{revtex}

\begin{document}

\preprint{TPR-00-04} 
\author{Ph. H\"{a}gler$^{1}$, R. Kirschner$^{2}$, A. 
Sch\"{a}fer$^{1}$, L. Szymanowski$^{1,3}$, O.V. Teryaev$^{1,4}$ \\   
$^{1}$Institut f\"{u}r Theoretische Physik, Universit\"{a}t Regensburg,
D-93040 Regensburg, Germany\\
$^{2}$Institut f\"{u}r Theoretische Physik, Universit\"{a}t Leipzig,
D-04109 Leipzig, Germany\\
$^{3}$Soltan Institute for Nuclear Studies,
Hoza 69, 00681 Warsaw, Poland\\
$^{4}$ Bogoliubov Laboratory of Theoretical Physics, 
JINR, 141980 Dubna, Russia
}  
\title{Heavy quark production as sensitive test for an improved
description of high energy hadron collisions}
  
\maketitle 
 
\begin{abstract}  
QCD dynamics at small quark and gluon momentum fractions or 
large total energy, which plays a major role for HERA, the Tevatron, RHIC and LHC physics, 
is still poorly understood. For one of the simplest processes, namely  
$b\bar{b}$ production, next-to-leading-order perturbation theory fails.  
We show that the combination of two recently developed theoretical concepts, the  
$k_\perp$-factorization and the next-to-leading-logarithmic-approximation  
BFKL vertex, 
 gives perfect agreement with data. One can therefore hope  
that these concepts provide a  
valuable foundation for the description of other high energy processes.

\end{abstract}

  
Existing QCD calculations describe
many high energy observables which involve partonic transverse
momentum rather poorly.
This is also true for the theoretically especially clean case of $b\bar{b}$ 
production, which was investigated experimentally at Fermilab \cite{Abe97}. 
Since central 
quark-antiquark production at $\sqrt{s}=1.8$ TeV is sensitive to  
very small gluon momentum fraction  
$x\thickapprox 10^{-2}-10^{-4}$, one probes the gluon content of the nucleon  
at small $x$, which is a central issue of current research.  We 
reconsider this process and combine as essential new ingredients 
the $k_{t}$-factorization scheme with the next-to-leading-logarithmic-approximation (NLLA) BFKL production vertex derived in \cite{FL96}. 
The $k_{t}$-factorization approach for the description  
of high energy processes \cite{CCH90,CE91,CC96,RSS99} differs strongly from the conventional NLO 
collinear approximation (e.g. \cite{CSS88}) because it takes the non-vanishing transverse  
momenta of the scattering partons into account. The usual gluon densities  
are replaced by unintegrated gluon distributions which depend on the  
transverse momentum $k_{t}$. These together  
with the $k_{t}$-factorization form a basis for a general calculation
scheme for high energy (i.e. small $x$). 
The standard collinear approximation has the advantage of being 
closely related to the operator product expansion. It is, however, 
only justified for the processes dominated by $x={\cal O}(1)$. In 
application to processes governed by small $x$ the 
$k_{t}$-factorization 
approach has the advantage that its approximations
correspond to the dominant kinematics. Essential small x 
contributions 
are included in the Born approximation which in the collinear approach  
are accounted for in higher orders only. This is well known from the 
case of structure functions where the DGLAP evolution is appropriate 
for $x={\cal O}(1)$ and the BFKL evolution for small $x$. 
 
While the $k_{t}$-factorization formalism is very attractive 
theoretically, its phenomenological usefullness has been mostly tested  
in the case of the structure function $F_{2}$ \cite{BE96,KMS97}. The 
NLLA BFKL vertices are just the ones needed to treat semi-hard central  
production at collider energies in this approach. 
 
In our  
calculation we use one particular element of the NLLA BFKL 
formalism \cite{FL96,F98}, namely the effective  
vertex for quark-antiquark production. Thus our calculation can be seen as a  
first phenomenological application of this vertex which decides
whether the NLLA BFKL formalism can be hoped to converge.  
  
One special aspect of the reaction we investigate is the possible loss of  
gauge invariance when a $q\overline{q}$ production vertex is incorporated  
into an amplitude with off-shell gluons. In the BFKL approach, however,  
gauge invariance is ensured automatically by the use of the just mentioned  
NLL effective vertex which is valid in quasi multi Regge kinematics  
(QMRK), i.e. when the $q$ and $\overline{q}$ have similar rapidities and  
form a cluster (in contrast to LLA, where the particles are produced with a  
large rapidity gap).  
 

\begin{figure}[h] 
 \centerline{\epsfig{file=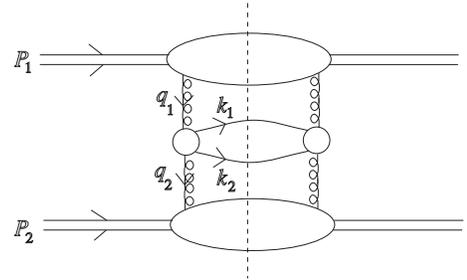,width=6cm}} 
 \caption{The basic diagram} 
 \label{cutdia} 
\end{figure} 
We begin with the following definition for the light cone coordinates 
and the momenta of the scattering hadrons in the c.m. frame  
\begin{eqnarray} 
&&k^{+}=k^{0}+k^{3},\,k^{-}=k^{0}-k^{3},\,k_{\perp  
}=(0,k^{1},k^{2},0)=(0,{\mathbf k},0).\nonumber\\ 
&&P_{1}^{+}=P_{2}^{-}=\sqrt{s},\;P_{1}^{-}=P_{2}^{+}=0,\; 
P_{1\perp }=P_{2\perp }=0.\nonumber 
\end{eqnarray} 
The Mandelstam variable $s$ is as usual the c.m. energy squared. 
As defined in fig. \ref{cutdia}, $q_{1}$ and $q_{2}$ are the momenta of the  
gluons and the on-shell quark and antiquark have momentum $k_{1}$ 
respectively $k_{2}$.   
In the high energy (large $s$) regime we have   
\begin{eqnarray*}  
&&k_{1}^{+}+k_{2}^{+} =q_{1}^{+}-q_{2}^{+}\approx q_{1}^{+}, \\  
&&k_{1}^{-}+k_{2}^{-} =q_{1}^{-}-q_{2}^{-}\approx -q_{2}^{-}, \\  
&&q_{1}^{2} \approx q_{1\perp }^{2}, 
q_{2}^{2} \approx q_{2\perp }^{2}.  
\end{eqnarray*}  
The longitudinal momentum fractions of the gluons are $%
x_{1}=q_{1}^{+}/P_{1}^{+}$, $x_{2}=-q_{2}^{-}/P_{2}^{-}$.

The cross section for heavy quark pair production in the $k_{t}$%
-factorization approach is then given by \cite{CCH90,CE91}  
  
\begin{eqnarray}  
\sigma&& _{P_{1}P_{2}\rightarrow q\overline{q}X} =\frac{1}{16(2\pi )^{4}}%
\int \frac{d^{3}k_{1}}{k_{1}^{+}}\frac{d^{3}k_{2}}{k_{2}^{+}}d^{2}q_{1\perp  
}d^{2}q_{2\perp } 
\nonumber\\ 
&&\delta ^{2}(q_{1\perp }-q_{2\perp }-k_{1\perp }-k_{2\perp })  
{\mathcal F}(x_{1},q_{1\perp }) 
\frac{1}{(q_{1\perp }^{2})^{2}} 
\nonumber \\  
&&\left\{ \frac{%
\psi ^{\dagger c_{2}c_{1}}\psi ^{c_{2}c_{1}}}{(N^{2}-1)^{2}}\right\} \frac{1%
}{(q_{2\perp }^{2})^{2}}{\mathcal F}(x_{2},q_{2\perp }).  \label{cs}  
\end{eqnarray}  
The factor $(N^{2}-1)^{2}$ reflects the projection on color singlet, where $%
N $ is the number of colors. The hard amplitude $\psi  
^{c_{2}c_{1}}(x_{1},x_{2},q_{1\perp },q_{2\perp },k_{1},k_{2})$ is  
calculable in perturbation theory, whereas the unintegrated gluon  
distribution ${\mathcal F}(x,q_{\perp })$ has to be measured or modelled. We  
choose the argument $\mu ^{2}$ of the strong coupling constant $\alpha  
_{S}(\mu ^{2})$ in the hard amplitude $\psi ^{c_{2}c_{1}}$ to be equal to $%
{\mathbf q}_{1}^{2}=-q_{1\perp }^{2}$ respectively ${\mathbf q}%
_{2}^{2}=-q_{2\perp }^{2}$ \cite{LRSS91}.  
 
\begin{figure}[h] 
 \centerline{\epsfig{file=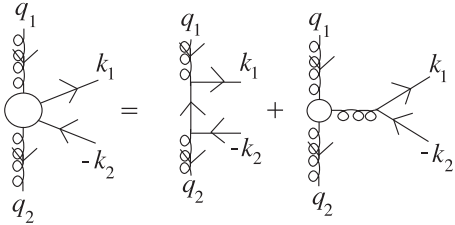,width=6cm}} 
 \caption{The effective vertex} 
 \label{flvertex} 
\end{figure}  
We generalize the  
results on the $q\bar{q}$ production vertex presented in \cite{FL96} for  
massless QCD in an obvious way in order to take the masses $m$ of the  
produced quarks into account. The resulting vertex $\Psi ^{c_{2}c_{1}}$ is  
given by a sum of two terms   
\begin{equation}  
\label{FLvertex} 
\Psi ^{c_{2}c_{1}}=-g^{2}\left(  
t^{c_{1}}t^{c_{2}}\,b(k_{1},k_{2})-t^{c_{2}}t^{c_{1}}\,b^{T}(k_{2},k_{1})%
\right) ,    
\end{equation} 
where $t^{c}$ are the colour group generators in the fundamental  
representation. The connection between $\psi ^{c_{2}c_{1}}$ in eq. (\ref{cs}%
) and $\Psi ^{c_{2}c_{1}}$ in eq. (\ref{FLvertex}) is given by   
\[  
\psi ^{c_{2}c_{1}}=\overline{\mbox{u}}(k_{1})\Psi ^{c_{2}c_{1}}\mbox{v}  
(k_{2}),  
\]  
with the on-shell quark and antiquark spinors u$(k)$ and v$(k)$. The  
expression for $b(k_{1},k_{2})$ is a sum of two terms   
\begin{equation} 
\label{vertex1}  
b(k_{1},k_{2})=\gamma ^{-}\frac{\not{q}_{1\perp }-\not{k}_{1\perp }-m}{%
(q_{1}-k_{1})^{2}-m^{2}}\gamma ^{+}-\frac{\gamma _{\beta }\Gamma  
^{+\,-\,\beta }(q_{2},q_{1})}{(k_{1}+k_{2})^{2}}\,,   
\end{equation} 
The first term on the r.h.s. of eq. (\ref{vertex1}) describes the production  
of a ${q\bar{q}}$ pair by means of usual vertices (see fig. \ref{flvertex}),  
the second term involves the light-cone projection of the effective vertex $%
\Gamma ^{+\,-\,\beta }(q_{2},q_{1}),$ which describes the transition of two $%
t$-channel gluons (reggeons) with momenta $q_{1}$ and $q_{2}$ to a gluon  
with momentum $k_{1}+k_{2}$   
\begin{eqnarray} 
\Gamma ^{+\,-\,\beta }(q_{2},q_{1})&=&2(q_{1}+q_{2})^{\beta  
}-2q_{1}^{+}n^{-\beta }-2q_{2}^{-}n^{+\beta } 
\nonumber \\  
&&-2t_{1}\,\frac{n^{-\beta }}{%
q_{1}^{-}-q_{2}^{-}}+2t_{2}\,\frac{n^{+\beta }}{q_{1}^{+}-q_{2}^{+}}, 
\label{Gam}   
\end{eqnarray} 
with $t_{1/2}=q_{1/2}^{2}$. This effective vertex differs from the usual  
triple-gluon vertex by the appearence of the last two terms. They are  
related to Feynman diagrams in which the ${q\bar{q}}$ pair is not produced  
by the $t$-channel gluons but in other ways. These two last terms in eq. (%
\ref{Gam}) are also required by gauge invariance,  
$\Gamma ^{+\,-\,\beta }(q_{2},q_{1})(q_{1}-q_{2})_{\beta }=0$.  
Another consequence of gauge invariance is the vanishing of the matrix  
element of the effective vertex $\Psi ^{c_{2}\,c_{1}}$ between on-mass-shell  
quark and antiquark states in the limit of small $q_{1\perp }$ or $q_{2\perp  
}$%
\[  
\overline{\mbox{u}}(k_{1})\Psi ^{c_{2}\,c_{1}}\mbox{v}(k_{2})\to 0\mbox{ \  
for }q_{1\perp }\mbox{ or }q_{2\perp }\to 0.  
\]  
The function $b^{T}(k_{2},k_{1})$ is very similar to (\ref{vertex1})   
\[  
b^{T}(k_{2},k_{1})=\gamma ^{+}\frac{\not{q}_{1\perp }-\not{k}_{2\perp }+m}{%
(q_{1}-k_{2})^{2}-m^{2}}\gamma ^{-}-\frac{\gamma _{\beta }\Gamma  
^{+\,-\,\beta }(q_{2},q_{1})}{(k_{1}+k_{2})^{2}}\;.  
\]  
 
 
The unintegrated gluon distribution is related to the standard gluon  
distribution by   
\[  
xg(x,{\mathbf q}^{2})=\int_{0}^{\infty }\frac{d{\mathbf 
k}^{2}}{{\mathbf k}^{2}}%
\Theta ({\mathbf q}^{2}-{\mathbf k}^{2}){\mathcal F}(x,{\mathbf k}).  
\]  
Taking the derivative of this expression makes it obvious that 
${\mathcal F}%
(x,\mathbf{k})$ includes the evolution of $xg(x,\mathbf{q}^{2})$, which is  
given by the BFKL and/or DGLAP equation. Since the unintegrated gluon  
distribution is not known at small $\mathbf{k}$, we write this equation as   
\begin{equation} 
\label{integrated}  
xg(x,{\mathbf q}^{2})=xg(x,q_{0}^{2})+\int_{q_{0}^{2}}^{\infty }\frac{d%
{\mathbf k}^{2}}{{\mathbf k}^{2}}\Theta ({\mathbf q}^{2}-{\mathbf k}^{2})%
{\mathcal F}(x,{\mathbf k}).   
\end{equation} 
This formula has been repeatedly used \cite{K85,RS94,CE91,RSS96} and  
introduces the a priori unknown initial scale $q_{0}$ and the initial gluon  
distribution $xg(x,q_{0}^{2})$. Following \cite{KMS97}, one may neglect the  
hard cross section dependence on $\mathbf{q}$ in the soft region 
$|{\mathbf q}%
|<q_{0}$, so that 
   
\begin{eqnarray}  
\frac{1}{q_{1\perp }^{2}}&&\left\{ \frac{\psi ^{\dagger c_{2}c_{1}}\psi  
^{c_{2}c_{1}}}{(N^{2}-1)^{2}}\right\} \frac{1}{q_{2\perp }^{2}} \equiv  
S(q_{1\perp },q_{2\perp })\to  
\nonumber\\ 
&&\;S(q_{1\perp },q_{2\perp })\Theta ({\mathbf q}%
_{1}^{2}-q_{0}^{2})\Theta ({\mathbf q}_{2}^{2}-q_{0}^{2})   
\nonumber \\  
+&&\;S(q_{1\perp },0)\Theta ({\mathbf q}_{1}^{2}-q_{0}^{2})\Theta (q_{0}^{2}-%
{\mathbf q}_{2}^{2})   
\nonumber \\  
+&&\;S(0,q_{2\perp })\Theta ({\mathbf q}_{2}^{2}-q_{0}^{2})\Theta (q_{0}^{2}-%
{\mathbf q}_{1}^{2})   
\nonumber \\  
+&&\;S(0,0)\Theta (q_{0}^{2}-{\mathbf q}_{1}^{2})\Theta 
(q_{0}^{2}-{\mathbf q}%
_{2}^{2}).  \label{prescript}  
\end{eqnarray}

Note that the very existence of the finite limit $q_{\perp }\rightarrow 0$  
follows from the decrease of the production amplitude due to gauge  
invariance. Substituting this formula in (\ref{cs}) using eq. (\ref  
{integrated}) one may easily perform the integration over $q_{\perp }$. As a  
result, $S(0,0)$ produces the standard expression of collinear factorization  
(ref. \cite{KMS97,RSS96}), while $S(q_{1\perp },0)$, $S(0,q_{2\perp })$  
correspond to the asymmetric configurations, where one of the gluons is  
described by the unintegrated distribution and the other by the integrated  
one. Here it is important to notice that when we insert (\ref{integrated}), (%
\ref{prescript}) in (\ref{cs}) the coupling constant $\alpha _{s}$ in the  
term proportional to $xg(x,q_{0}^{2})$ is taken to be $\alpha _{s}(q_{0}^{2})  
$.

\begin{figure}[h] 
 \centerline{\epsfig{file=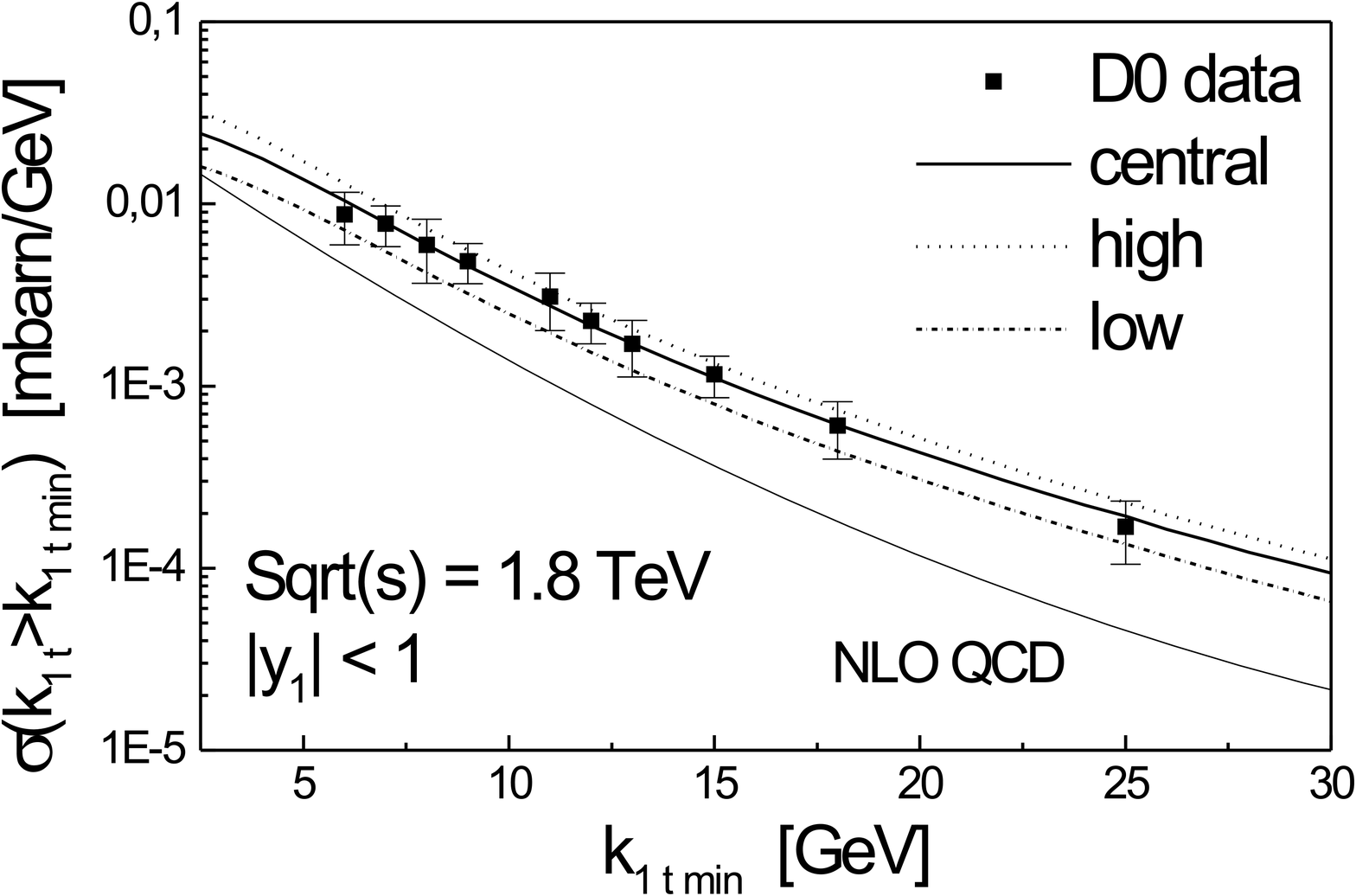,width=6cm}} 
\caption{The calculated $b$ cross section in comparison to
experimental data 
from the D0 Collaboration and the NLO QCD result 
with MRSR2 structure functions and $m_b=4.75$ GeV from [15]}
 \label{singleb} 
\end{figure} 
In all  
our numerical calculations we used for the unintegrated gluon  
distribution ${\mathcal F} (x,\mathbf{k})$ the code 
by Kwiecinski, Martin and Sta\'{s}to \cite{KMS97}, 
because they use a combination of DGLAP and BFKL equations which 
governs simultaneously the evolution in $Q^{2}$ and $x$. They obtain 
an excellent description of $F_{2}(x,Q^{2})$ in a very large 
$x$-$Q^{2}$-window. According to our knowledge this is the only 
unintegrated gluon distribution which has given such a satisfactory 
result, which justifies our choice. 
As in the case of the usual gluon distribution function one has to 
choose an initial scale and an initial distribution function which in 
the case of \cite{KMS97} are given by 
\begin{eqnarray}  
q_{0}^{2} &=&1\mbox{ GeV}^{2},\;\, 
xg(x,q_{0}^{2}) =1.57(1-x)^{2.5}.  \label{inivals}  
\end{eqnarray}  
We use these values, which are fixed by the fit to $F_{2}(x,Q^{2})$, 
in our calculation. 
 
 
We consider the  
production of $b\overline{b}$-pairs. For the computation we use eqs. (\ref  
{cs}) and (\ref{prescript}) with the unintegrated gluon distribution from   
\cite{KMS97} and the corresponding values (\ref{inivals}). The 
rapidities and 
the transverse masses  of  
the produced quark and antiquark are defined by   
\[  
y_{1/2}=\frac{1}{2}\ln (\frac{k_{1/2}^{+}}{k_{1/2}^{-}}),\;\, 
m_{1/2\perp }=\sqrt{m^{2}-k_{1/2\perp }^{2}}.  
\]  
The Bjorken-variables of the gluons can then be written as   
\begin{eqnarray*}  
x_{1} &=&\frac{1}{\sqrt{s}}(m_{1\perp }e^{y_{1}}+m_{2\perp }e^{y_{2}}),\\ 
x_{2} &=&\frac{1}{\sqrt{s}}(m_{1\perp }e^{-y_{1}}+m_{2\perp }e^{-y_{2}}).  
\end{eqnarray*}  
In fig. \ref  
{singleb} we show our results for inclusive $b$ production, together with  
experimental results measured by the D0 Collaboration \cite{Abb99} (see  
Table II) in$\sqrt{s}=1.8$ TeV $p\overline{p}$ collisions. We obtain this  
cross section by integrating out all antibottom variables in eq. (\ref{cs}).  
\begin{figure}[h] 
 \centerline{\epsfig{file=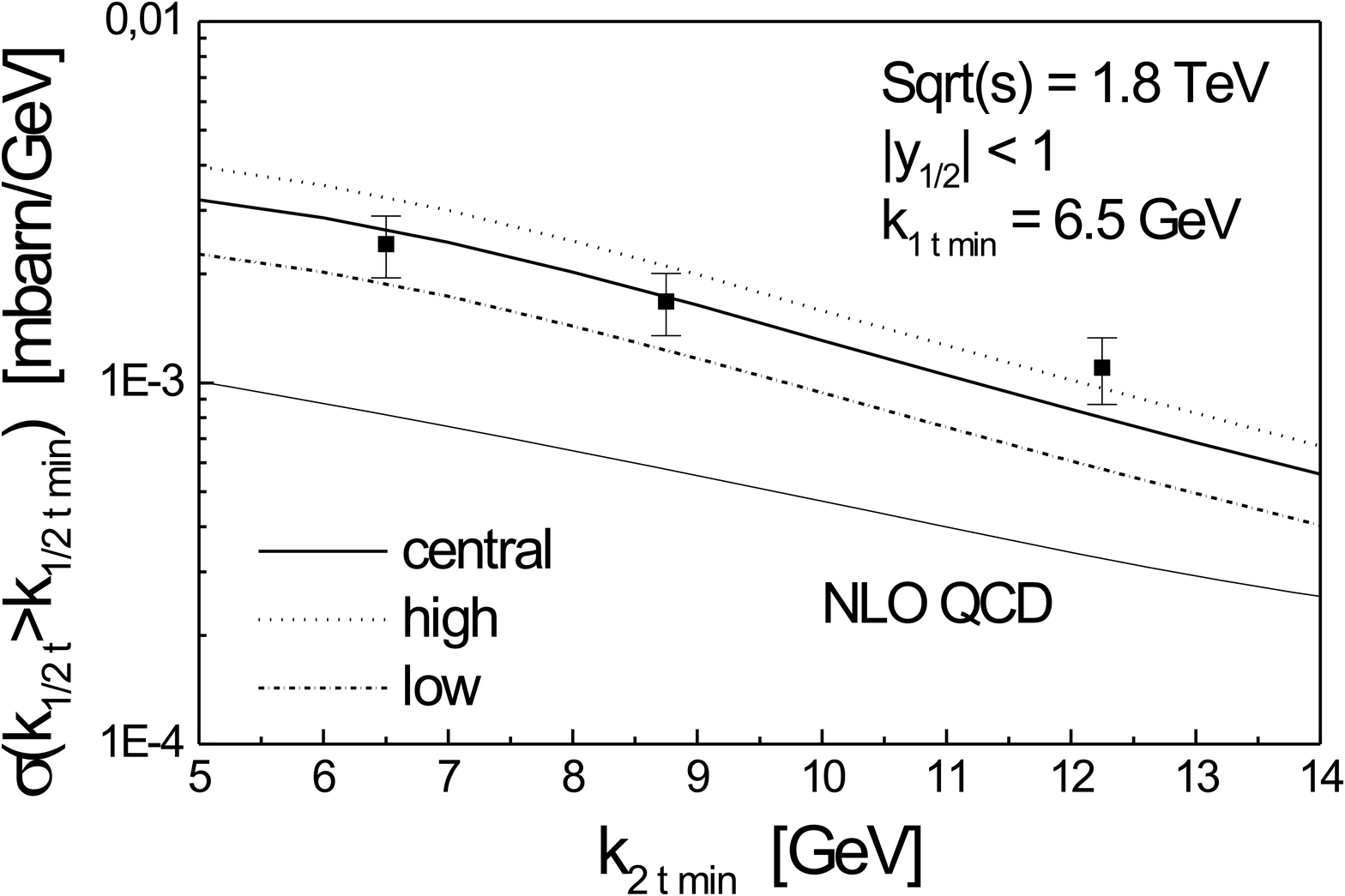,width=6cm}} 
 \caption{The result for the semi  
 differential $b\overline{b}$ cross section  
 at $k_{1\perp \min }=6.5$ GeV, compared to CDF data and the NLO QCD
result with MRSD0 structure functions and $m_b=4.75$ GeV from [1]} 
 \label{bbbar1} 
\end{figure}  
The variable $k_{1\perp \min }$ is the lower integration cut on the  
transverse momentum of the produced $b$ quark.  
To get an indication of the theoretical uncertainties apart from
higher 
order contributions which are not available at the moment we proceed in a 
similar way as the authors of ref. \cite{Abe97} and present our 
calculations for three different choices of $\Lambda _{\mbox{QCD}}$ and the  
bottom quark mass 
\begin{eqnarray*}  
\mbox{high } &:&\mbox{ }\Lambda ^{(5)}=180\mbox{ MeV, }m_{b}=4.5\mbox{ GeV,}  
\\  
\mbox{central } &:&\mbox{ }\Lambda ^{(5)}=150\mbox{ MeV, }m_{b}=4.7\mbox{  
GeV,} \\  
\mbox{low } &:&\mbox{ }\Lambda ^{(5)}=100\mbox{ MeV, }m_{b}=4.9\mbox{ GeV,}  
\end{eqnarray*}  
Our result is in  
very good quantitative agreement with data over the whole range of $%
k_{1\perp \min }$. The corresponding central QCD NLO calculation has a 
similar shape, but is about a 
factor of $2-3$ smaller than our central result (see for example fig. 11 in \cite{N00}). 
\begin{figure}[h] 
 \centerline{\epsfig{file=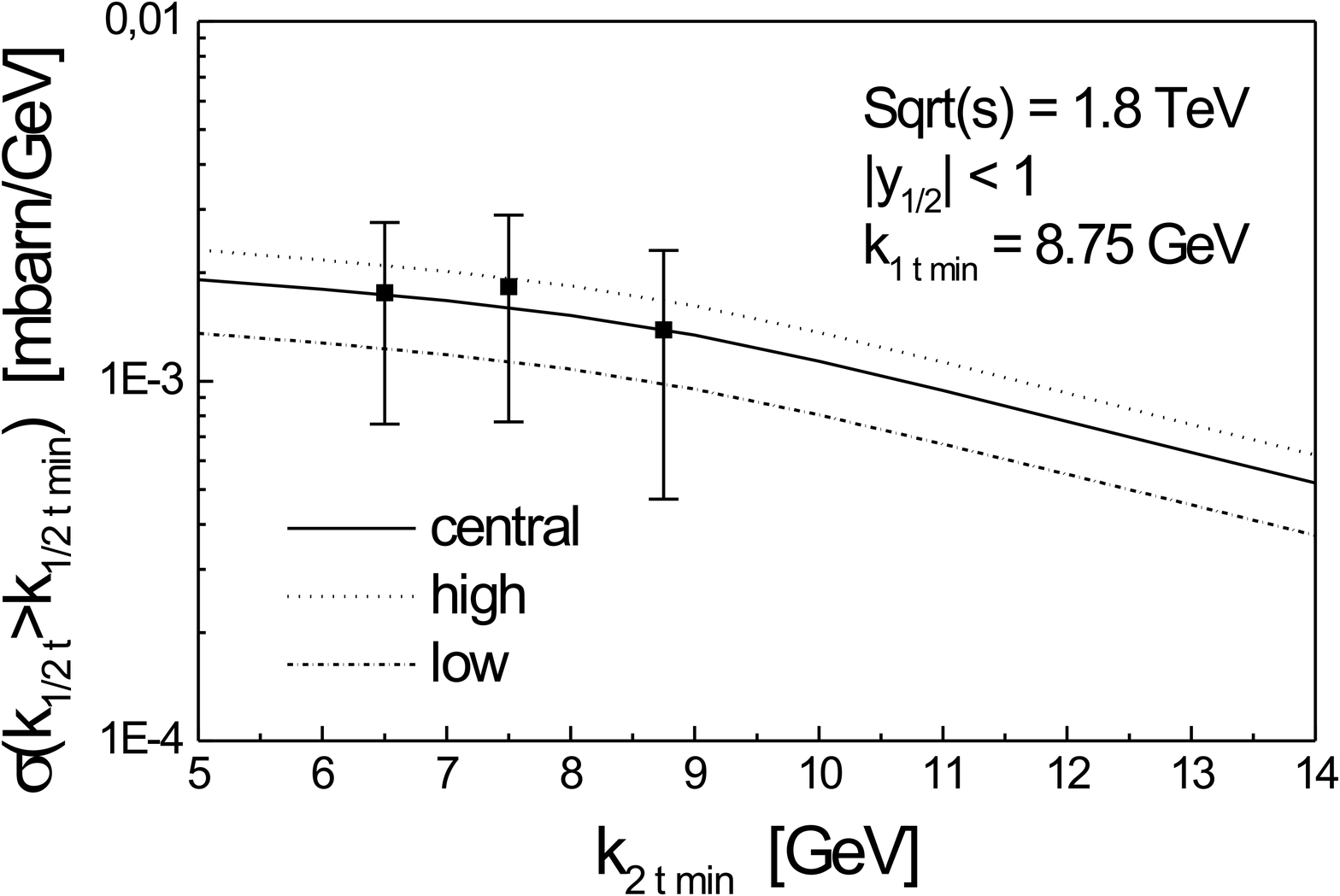,width=6cm}} 
 \caption{The result for the semi  
 differential $b\overline{b}$ cross section 
 at $k_{1\perp \min }=8.75$ GeV, compared to CDF data} 
 \label{bbbar2} 
\end{figure}  
We now turn to $b\overline{b}$ correlations in $\sqrt{s}=1.8$ TeV $p%
\overline{p}$ collisions, which have been measured by the CDF collaboration  
at Fermilab \cite{Abe97}. The correlations of the quark and antiquark give  
an insight into the dynamics of the production mechanism and are important  
in order to study the limits of the collinear ($k_{1\perp }=-k_{2\perp })$  
LO QCD approximation. We present a comparison between our results and the  
experimental data in fig. \ref{bbbar1} and fig. \ref{bbbar2}. The data  
points and uncertainties were taken from \cite{Abe97,Abe95}. 
We find  
good agreement with experiment for both $k_{1\perp \min }=6.5$ GeV (fig. \ref  
{bbbar1}) and $k_{1\perp \min }=8.75$ GeV (fig. \ref{bbbar2}).   
In this  
case QCD NLO calculations underestimate the measured cross section 
roughly by factor of $3$ (compare with fig.6 in \cite{Abe97}). 
\begin{figure}[h] 
 \centerline{\epsfig{file=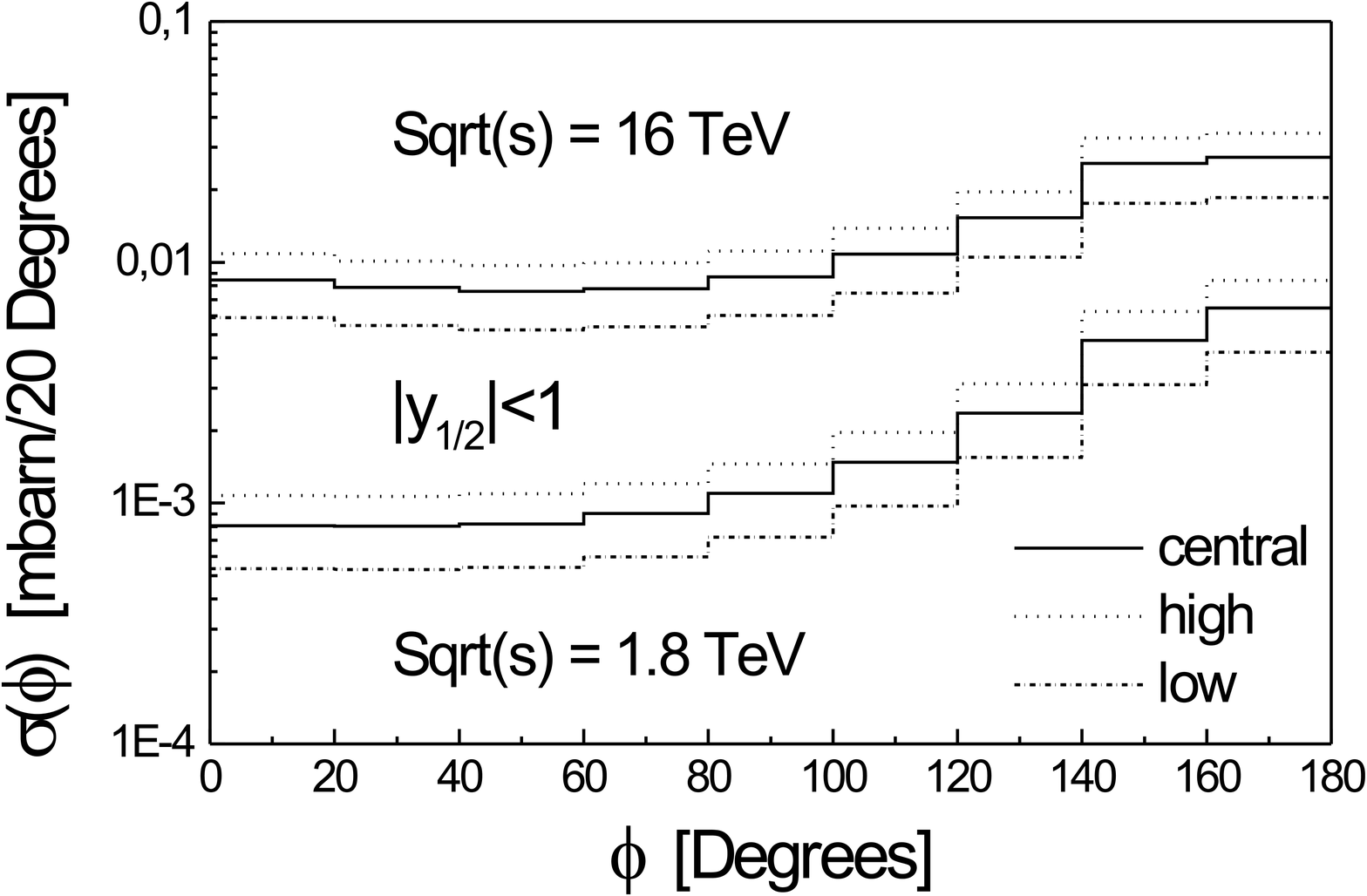,width=6cm}} 
 \caption{$\phi $ distribution of $b%
\overline{b}$ hadroproduction} 
 \label{bbbar_phi} 
\end{figure} 
 
An interesting  
parameter concerning the correlation is the opening angle $\phi $  
between the momentum vectors of the produced quarks in the plane transverse  
to the beam axis. Our predictions for the corresponding differential cross  
sections at Fermilab and LHC energies are shown in fig. \ref{bbbar_phi}. As  
expected we find a peak at $\phi =180{{}^{\circ }}$ which shows the  
dominance of the collinear part.  
 
\begin{figure}[h] 
 \centerline{\epsfig{file=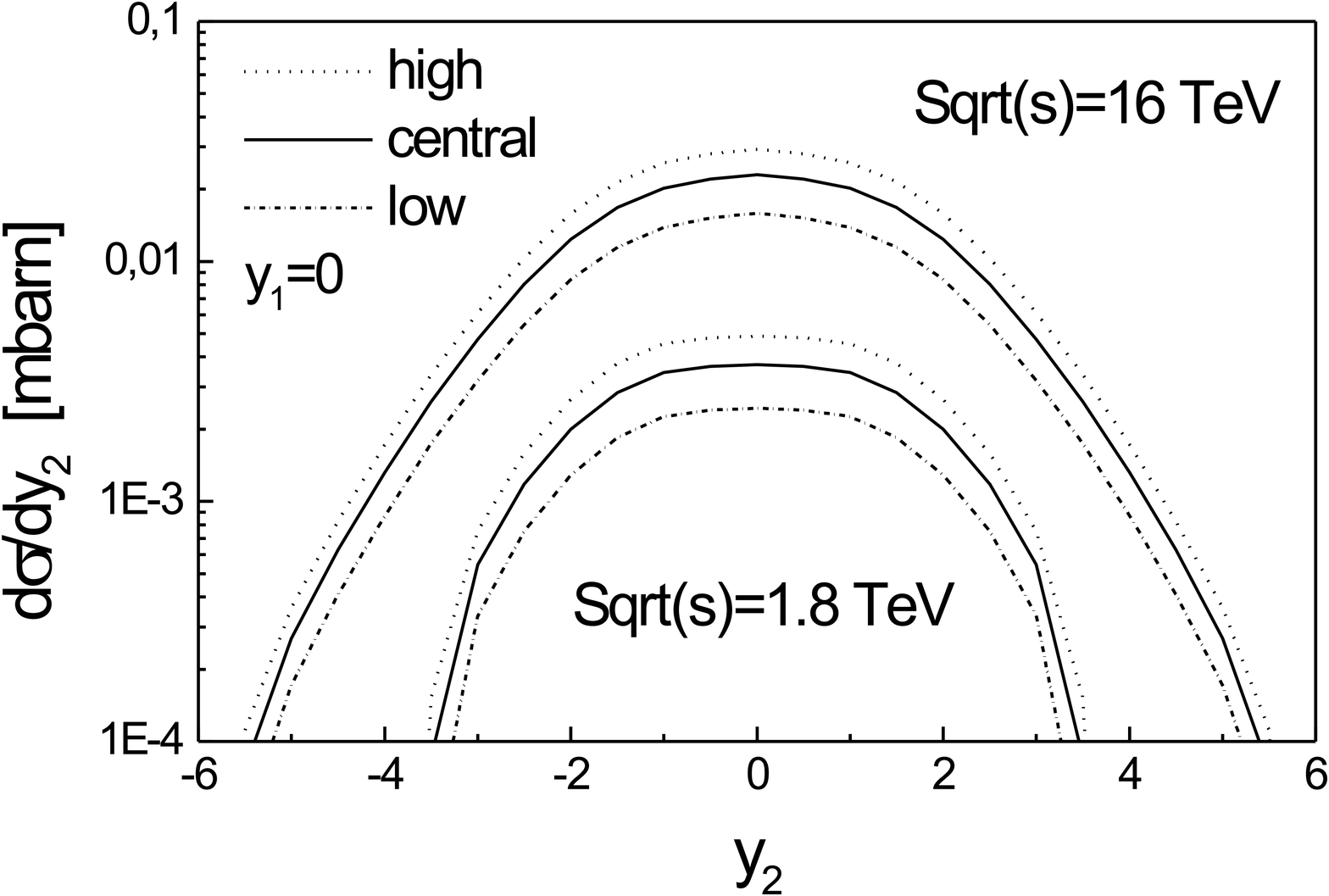,width=6cm}} 
 \caption{Rapidity distributions of $b\overline{b}$ hadroproduction} 
 \label{bbbar_y} 
\end{figure} 
 
 Additionally we present our predictions for rapidity distributions of the $%
\overline{b}$ for the rapidity of the $b$ being $0$ and $\sqrt{s}=1.8$
TeV respectively $\sqrt{s}=16$ TeV in fig. \ref{bbbar_y}.  
Our cross section for $\sqrt{s}=1.8$ TeV at $y_{2}=0$ is about a 
factor of $3$ larger than the corresponding QCD NLO result from \cite{MNR91}. 
 
 
Let us conclude. 
We have studied quark-antiquark hadroproduction within the $%
k_{t}$-factorization approach using an unintegrated gluon distribution and  
a specific effective BFKL vertex for $q\overline{q}$ production.  We found very good  
agreement with experiment for both single $b$ production and $b\overline{%
b}$ correlations at $\sqrt{s}=1.8$ TeV.  
Our approach leads to nontrivial $b\overline{b}$ correlations already at LO  
perturbation theory, whereas traditional collinear factorization gives them  
only at NLO and beyond. In contrast, the available NLO caculations \cite{MNR91} are not  
in agreement with the Tevatron data we compare with \cite{Abe97,Abe95,Abb99}.  
 
  
Our results show that at least those features of the effective $q\overline{q}  
$ vertex which we tested provide a substantial improvement with respect to  
the standard collinear treatment.  
  
If further tests of other observables should be equally successfull, the NLL  
BFKL vertices will also allow for a much improved description of many  
processes to be studied at RHIC and LHC.  
  
We thank J. Kwiecinski, A. Martin and A. Sta\'{s}to for supplying us with  
their code for the calculation of the unintegrated gluon distribution.  
  
L. S. would like to acknowledge the support by the DFG.

\end{document}